\documentclass[twocolumn]{article}
\usepackage[letterpaper, margin=0.8in]{geometry}
\usepackage[noblocks]{authblk}
\usepackage{latexsym}
\usepackage{graphics}
\usepackage{multirow}
\usepackage{textcomp}
\usepackage{siunitx}
\usepackage{amsmath}
\usepackage{xcolor}
\usepackage{dsfont}

\newcommand{\ket}[1]{\ensuremath{\left| {#1} \right>}}

\newcommand{\nbar}{\bar{n}}

\newcommand{\Be}{\ensuremath{{^9}{\rm Be}^{+} \,}}
\newcommand{\BeNoSpace}{\ensuremath{{^9}{\rm Be}^{+}}}
\newcommand{\Mg}{\ensuremath{{^{25}}{\rm Mg}^{+} \,}}
\newcommand{\MgNoSpace}{\ensuremath{{^{25}}{\rm Mg}^{+}}}
\newcommand{\Al}{\ensuremath{{^{27}}{\rm Al}^{+} \,}}
\newcommand{\Ca}{\ensuremath{{^{40}}{\rm Ca}^{+} \,}}

\usepackage[normalem]{ulem}

\newcommand{\create}{\ensuremath{{\,\hat{a}^{\dagger}}}}
\newcommand{\destroy}{\ensuremath{{\,\hat{a}}}}

\begin{document}

	\title{Quantum logic spectroscopy with ions in thermal motion}
	
	\author[1,2]{D. Kienzler\footnote{Authors D.K. and Y.W. have contributed equally to this work.}\footnote{Current address: ETH Zurich, 
					Otto-Stern-Weg 1, HPF E10,
					8093 Zurich, Switzerland}}	
	\author[1,2]{Y. Wan\textsuperscript{*}}
	\author[1,2]{S. D. Erickson}
	\author[1,2]{J. J. Wu}
	\author[1,2]{A. C. Wilson}
	\author[1,2,3]{D. J. Wineland}
	\author[1,2]{D. Leibfried\footnote{Corresponding author, email: \\ dietrich.leibfried@nist.gov}}
	\affil[1]{National Institute of Standards and Technology,
	Time and Frequency Division 688, 325 Broadway, Boulder, CO 80305, USA}
	\affil[2]{Department of Physics, University of Colorado, Boulder, CO 80305, USA}
	\affil[3]{University of Oregon, Department of Physics, Eugene, OR 97403, USA}

	\date{}
	
	\maketitle
	
	\begin{abstract}
		A mixed-species geometric phase gate has been proposed for implementing quantum logic spectroscopy on trapped ions that combines probe and information transfer from the spectroscopy to the logic ion in a single pulse. We experimentally realize this method, show how it can be applied as a technique for identifying transitions in currently intractable atoms or molecules, demonstrate its reduced temperature sensitivity, and observe quantum-enhanced frequency sensitivity when it is applied to multi-ion chains. Potential applications include improved readout of trapped-ion clocks and simplified error syndrome measurements for quantum error correction.
	\end{abstract}
	
	\section{Introduction}
	Quantum logic spectroscopy (QLS) can be used for internal-state preparation and readout ofatomic and molecular ion species that lack a suitable electronic level structure to directly implement these tasks \cite{05Schmidt,08RosenbandShort,16Wolf,17Chou}. In principle, through the use of a ``logic ion'' (LI) and its motional coupling to a co-trapped ``spectroscopy ion'' (SI), QLS allows control over any ion species. The traditional QLS protocol, as presented in \cite{05Schmidt}, has two main limitations. First, it requires the ions to be cooled to near the motional ground state. Second, its readout efficiency scales poorly with the number of SIs, which could pose an obstacle to achieving the improved stability that could come from scaling quantum-logic-enabled atomic clocks to multiple ions \cite{16Schulte}. Methods have been developed to mitigate these effects using repetitive quantum nondemolition (QND) measurements \cite{07Koelemeij2,07Hume,11Hume}. However, applying them might not be possible due to unsuitable electronic structure, and repetitive measurements will decrease the duty cycle of the spectroscopy probe.
	
	Here we demonstrate a QLS method proposed in \cite{16Leibfried} based on a geometric phase gate that is often used in quantum information processing for multi-qubit entangling gates. This type of geometric phase gate has previously been used on a mixed-species ion pair to implement quantum logic readout with reduced temperature sensitivity as part of a controlled-NOT operation \cite{15Tan}. In that experiment, the interrogation and detection of the SI were accomplished with separate laser operations. 
	Here we explore a technique, using only a M\o lmer-S\o rensen (MS) interaction \cite{99Sorensen,99Solano,00Sorensen1,00Milburn2,03LeibfriedShort}, to simultaneously implement both the spectroscopy operation and transfer of SI state information to the LI for readout. This reduces temperature sensitivity compared to traditional QLS \cite{09Kirchmair}. Additionally, the technique can be applied to multiple SIs, for which it can exhibit Heisenberg-limited spectroscopic sensitivity \cite{92Braunstein,96Bollinger}. Assuming the LI and SI are singly charged, the coupling strength will depend on the mass ratio of the ions, with the strongest coupling for equal masses.  In multi-species experiments done so far, quantum logic spectroscopy has been accomplished with mass ratios of up to approximately four. Since state control of ions with masses from 9~u (\BeNoSpace) up to 176~u ($^{176}$Yt$^+$) has been achieved, it should be possible to perform quantum logic spectroscopy with singly charged ions having masses up to about $4 \times 176$~u.
	While our technique shares some features with previous work, which implemented Heisenberg-limited Ramsey \cite{04Leibfried,11Monz} and Rabi spectroscopy \cite{18Shaniv}, it extends metrology that takes advantage of entanglement to a wider range and number of spectroscopy ions \cite{16Leibfried}.
	
	\section{Description of the method}
	\subsection{Basic operations}
	Modeling ions as effective two-level (spin-1/2 with eigenstates \ket{\uparrow} and \ket{\downarrow}) systems with coupling through a shared harmonic oscillator normal mode of motion with frequency $\omega_m$ and eigenstates $\ket{n}$, we require interactions that can drive motional sideband transitions $\ket{\downarrow}\ket{n}\leftrightarrow\ket{\uparrow}\ket{n\pm1}$ with ``+" denoting a ``blue" sideband (BSB) and ``$-$" a ``red" sideband (RSB). Relative to the spin (``carrier'') transition, the detunings of the RSB and BSB are $\delta = \pm (\omega_m + \delta_{MS})$, where $\delta_{MS}$ serves to implement the MS interaction. Such interactions can be implemented with laser fields \cite{08Blatt} and microwave fields \cite{09Johanning,11Ospelkaus} and have been demonstrated on mixed-species pairs and triads \cite{15Ballance,15Tan,18Negnevitsky}. For simplicity we initially assume equal sideband Rabi frequencies $\Omega_{sb}$. If the BSB and RSB are applied simultaneously for $t_{MS} = 4\pi/\delta_{MS}$ (twice the duration of a typical MS entangling operation $U_{MS}$), and with detuning $\delta_{MS}=4\Omega_{sb}$, for an even number of ions $K$, the system undergoes (up to a global phase factor) a complete spin-flip
		\begin{align}
		U_{MS}^2 = \sigma_x^{(1)}\otimes\sigma_x^{(2)}\otimes ...\otimes \sigma_x^{(K)}.
		\end{align}
	Here, $\sigma_j^{(i)}$ are the Pauli operators $(j\in \lbrace x, y, z\rbrace)$ acting on ion $i$. We will call this operation a ``mutually-controlled multi-flip" (shortened to ``multi-flip" in the following.) However, if $K$ is odd, the interaction does not result in a spin flip and instead the system only accumulates a global phase. See Appendix \ref{sec:MS} for the full MS Hamiltonian and details of the implementation.
	
	Now, let us consider two species of ions (LIs and SIs) with pseudo-spin states labeled $\lbrace\ket{\uparrow_{\mathrm{LI}}},\ket{\downarrow_{\mathrm{LI}}}\rbrace$ and $\lbrace\ket{\uparrow_{\mathrm{SI}}},\ket{\downarrow_{\mathrm{SI}}}\rbrace$ driven by independent RSBs and BSBs but with the same $\Omega_{sb}$. Let the detunings of the LI sidebands be $\delta_{LI}=\pm (\omega_m + \delta_{MS})$ as above. The SI sidebands have an additional detuning so that $\delta_{SI} = \pm(\omega_m+\delta_{MS}) + \delta_S$. In this scheme, $\delta_S$ acts as a ``switch" to turn on/off the SI's participation in the interaction. The SI resonance frequency, which would initially be unknown in a general spectroscopy experiment, can also be located by scanning $\delta_S$.
	
	For $M$ SIs and $N$ LIs, if the BSBs and RSBs for both species are applied simultaneously and on resonance ($\delta_s = 0$), all ions will undergo a complete spin-flip as described above, provided $M+N$ is even. Far off resonance ($\delta_s \gg\Omega_{sb}$), only the parity of $N$ matters, with the LIs undergoing the interaction and the SIs unaffected. For example, if $N$ is even, the interaction that occurs is described by
		\begin{align}
		\mathbf{I}_{SI}\otimes\sigma_x^{(1)}\otimes\sigma_x^{(2)}\otimes ...\otimes \sigma_x^{(N)}.
		\end{align}
	where $\mathbf{I}_{SI}$ is the identity operator acting on all $M$ SIs and $\sigma_x^{(i)}$ acts on LI $i\in {1,...,N}$. In this paper, we consider the specific cases of $N=1$ and $M$ odd, or both $M$ and $N$ even.
	
	For $N=1$ and the SIs far off resonance, the effect on the single LI is just a phase factor $e^{i\pi / 4}$. For intermediate detuning, $\delta_S\simeq\Omega_{sb}$, the dynamics are complicated, leading to substantial populations in several basis states. The probability of the LI flipping as a function of $\delta_S$ (i.e. the spectral line-shape) is not straightforward to express analytically and depends on the temperature of the motional mode, but qualitatively the result on the logic ion is that it continuously transfers from no flip to a full flip as $\delta_S$ is swept from far detuned to on resonance.
	
	For $M$ and $N$ even, the LIs flip with both the SIs on resonance and far off resonance, since both $M+N$ and $N$ are even. Other states are populated for an intermediate detuning $\delta_S\simeq \Omega_{sb}$ leading to a characteristic, temperature-dependent line-shape.
	
	In all cases, the multi-flip retains the temperature insensitivity of the MS interaction in the Lamb-Dicke (LD) regime \cite{00Sorensen1}, so it can be used as a robust combined spectroscopy probe and readout method. Moreover, the center frequency of the resonance does not depend on the initial states of the two species (though the line-shape and line-width do). Hence when searching for the approximate location of an unknown transition, it is not necessary to prepare the SIs in a pure state \footnote{It is however necessary to prepare the SI state to be within the spectroscopy manifold (\ket{\uparrow},\ket{\downarrow}).}.
	As described so far, the multi-flip is analogous to Rabi spectroscopy and ideally achieves a Fourier-limited frequency resolution of $\sim1/t_{\rm MS}$ when scanning the detuning $\delta_S$. The characteristics of Ramsey spectroscopy appear when the MS interaction is applied twice with a duration of $t_{\rm MS}/2$ (effective $\pi/2$-pulses) before and after a period $T_{\rm R}$ of free evolution. Analogous to traditional Ramsey spectroscopy, this sequence ideally achieves Fourier-limited frequency resolution ($1/T_{\rm{R}}$ for a single SI) if the effective $\pi/2$-pulses are much shorter than the Ramsey free evolution time, i.e.\ $t_{\rm{MS}} \ll T_{\rm{R}}$. As discussed in the next section, for more SIs driven on resonance, $M$-partite entangled ``Schr{\"o}dinger-cat spin states'' \cite{00Sackett,05Leibfried,11Monz} evolve during $T_{\rm R}$ to ideally achieve Heisenberg-limited resolution.
	
	\subsection{Heisenberg scaling}
	In the Ramsey version of the multi-flip, the SIs evolve in a maximally entangled state, resulting in increased frequency sensitivity, i.e.\ providing Heisenberg scaling with the number of SIs \cite{96Bollinger,04Leibfried,11Monz,16Leibfried,18Shaniv}.
	For example, if an ensemble of $N$ SIs and $M$ LIs ($N,M$ even) is prepared in the spin-up state $\ket{\uparrow _{1},\uparrow _{2}, ... \uparrow _{N+M}}$, the first Ramsey pulse will result in a ``Schr\"odinger-cat spin state'' for the SIs and LIs. Its evolution is then mapped onto the LIs with the second Ramsey pulse.
	The LIs being part of the entangled state does not add frequency sensitivity. In principle a single LI would suffice, although a larger number of LIs could result in a better detection signal (for example, with three or more LIs one can perform a majority vote to decide whether the SIs have flipped or not) and also improve the initial cooling of the mixed-species ion chain. For best performance and the simplest implementation, the mode of motion being driven in the MS interaction should ideally have the same mode amplitudes for all ions of the same species. For long ion chains this could be approximately achieved by using the in-phase mode of motion (see Appendix \ref{sec:normal_mode_amplitudes}).
	
	\subsection{Influence of LI properties on the protocol}
	We note that by symmetry between the LIs and SIs in this protocol, any detuning of the LIs from resonance will shift the line-center in Rabi spectroscopy and the fringe pattern in Ramsey spectroscopy. In addition, any decoherence of the SIs or LIs will reduce the contrast of the signal. 
	
	To reduce the impact of frequency instability of the LI, it is advantageous to choose an LI with a low absolute frequency transition, for example a hyperfine transition ($\sim$1 GHz), to pair with an SI with an optical clock transition ($\sim$1 PHz). In this case, the fractional stability of the LIs can be $\sim10^6$ worse than that of the SIs, given by the frequency ratio of the two transitions.  When applying the technique to state-of-the-art optical atomic ion clocks with fractional frequency uncertainties at the $\sim10^{-18}$ level, this requires $\sim$1 GHz LI transitions to have frequency uncertainty of about $\sim10^{-12}$. This stringent requirement can be relaxed by deploying dynamic decoupling sequences, where suitable $\pi$-pulses on the LIs only refocus their phase during free-precession periods. This can reduce the contribution of the LIs to the accumulated phase by orders of magnitude.  Shifts in the clock transition due to LI driving fields can be suppressed by techniques similar to hyper-Ramsey and auto-balanced spectroscopy \cite{10Yudin, 12Huntemann, 18Sanner}. Finally, the dependence of the LI's spin population on both the LI and SI can also be used to implement ``designer atoms'' or an ``atomic combination clock'' using two clock species \cite{06Roos,18Akerman}, where one of the clock species does not require direct readout.
	
	\section{Implementation and Results}
	\subsection{Basic method}\label{sec:Implementation of basic method}
	We demonstrate the basic features of this method on a mixed-species ion pair of electronic ground-state hyperfine qubits composed of one SI (\Mg with the two-level system defined as $\ket{\uparrow_{\rm{Mg}}} = \ket{3, 1}$ and $\ket{\downarrow_{\rm{Mg}}} = \ket{2, 0}$, with a frequency splitting $\omega_{\mathrm{Mg}} \approx 2\pi \times \SI{1.7632}{\giga\hertz}$ at the applied magnetic field of $B\approx\SI{11.9}{\milli\tesla}$) and one LI (\Be with $\ket{\uparrow_{\rm{Be}}} = \ket{1, 1}$, $\ket{\downarrow_{\rm{Be}}} = \ket{2, 0}$, with a frequency splitting $\omega_{\mathrm{Be}} \approx 2\pi \times \SI{1.2075}{\giga\hertz}$ at the same magnetic field). The MS interaction is implemented on the in-phase mode of motion along the (axial) direction defined by the line connecting the equilibrium positions of both ions.
	We perform Rabi- and Ramsey-type experiments with both axial modes of motion initially cooled to the ground-state, while the LI is prepared in \ket{\uparrow} and the SI in either \ket{\uparrow} or \ket{\downarrow}. We repeat this procedure for a range of values of $\delta_{\rm{S}}$ and detect both the LI and SI spin states (
	see Appendix \ref{sec:state_prep} for details on cooling, state preparation, state detection and the MS interaction). The LI is driven on resonance throughout the entire series of experiments. 
	
\begin{figure}[ht!]
		\resizebox{0.48 \textwidth}{!}{
			\includegraphics{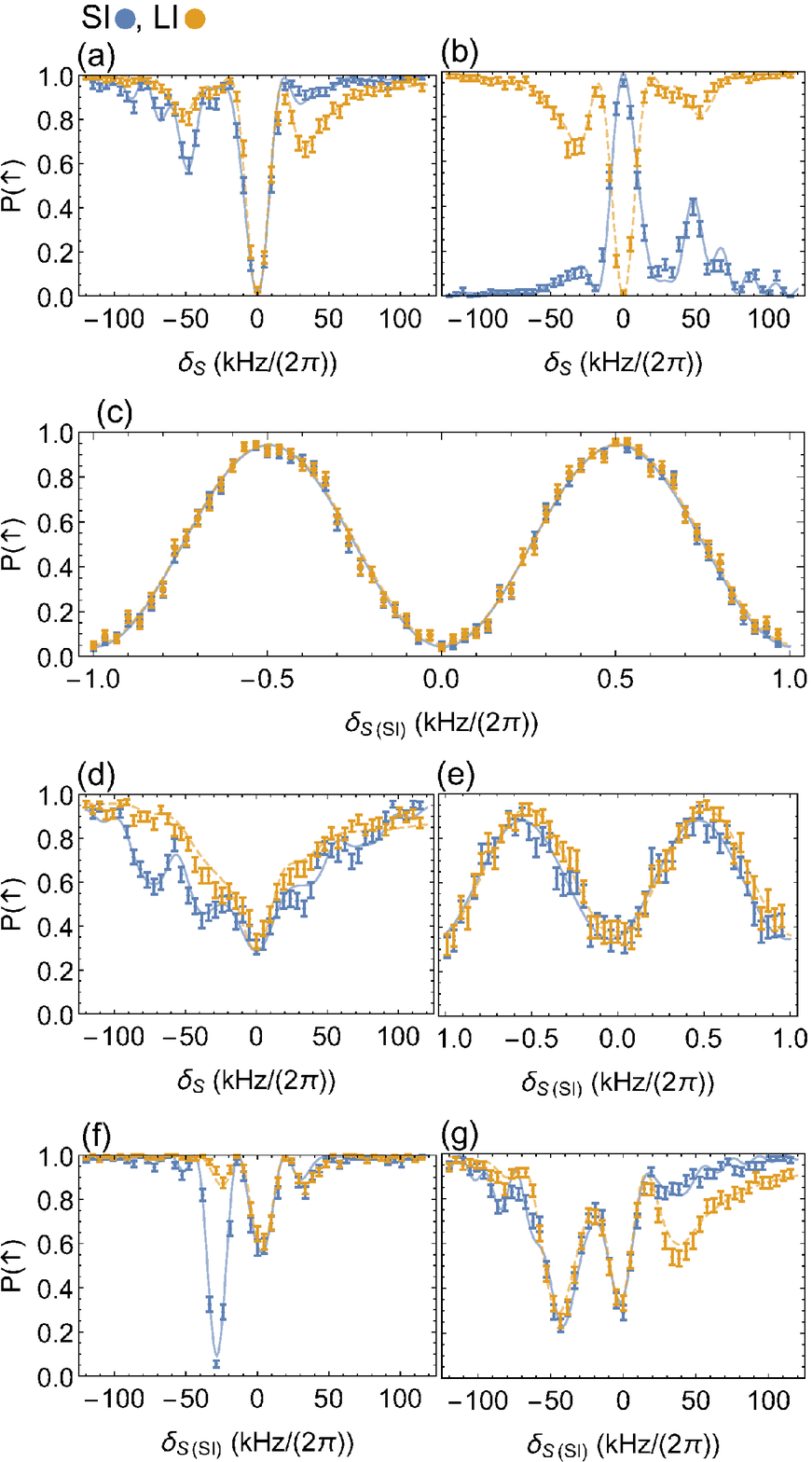}
		}
		\caption{
			Spin populations for the LI-SI pair (SI: blue, LI: orange). Fits are displayed in matching colors with solid/dashed curves for fits to SI/LI data. (a) Rabi spectroscopy with the SI's spin state initialized to \ket{\uparrow_{\rm{SI}}}, 
			(b) Rabi spectroscopy with the SI's spin state initialized to \ket{\downarrow_{\rm{SI}}}. 
			(c) Ramsey spectroscopy with a contrast of 0.89(1) (LI populations) and a phase shift of $\Delta\phi=0.01(1)$ between LI and SI sinusoids. (d) Rabi and (e) Ramsey spectroscopy with Doppler cooling only; Fits to Ramsey data give a contrast 0.57(2) (LI populations) and a phase shift $\Delta\phi=0.21(4)$ between LI and SI sinusoids. Rabi spectroscopy with (f) $\Omega_{\rm{SI}}= \Omega_{\rm{MS}}/2$ and (g) $\Omega_{\rm{SI}}= 2\, \Omega_{\rm{MS}}$. 
			In all figures, error bars (one sigma) are calculated assuming only quantum projection noise \cite{93Itano}. Each data point represents the average outcome of 200 repetitions of the experiment (100 for figure e).}
		\label{fig:two-ion}
	\end{figure}

	As predicted for the multi-flip protocol, both the LI and SI show a clear simultaneous resonance at $\delta_{\rm{S}}=0$ (Fig. \ref{fig:two-ion} (a), (b)). The asymmetries visible at larger detuning arise from the red-/blue-detuned light fields of the MS interaction becoming resonant with red or blue sideband transitions when $\delta_{\rm{S}}=\pm\delta_{\rm{MS}}$. For large values of $\delta_{\rm{S}}$, the spin states of both ions remain unchanged.
	To fit the full spectrum of these Rabi-type experiments we numerically simulate the evolution of the full system Hamiltonian without making the LD approximation, truncating the harmonic oscillator Hilbert space at $n=10$. We fit the resulting spin populations to the data. We use the parameter $c$ to define $\Omega_{\rm{MS(LI)}} = c \, \Omega_{\rm{MS}}$ and  $\Omega_{\rm{MS (SI)}} = \Omega_{\rm{MS}}/c$ and optimize by varying $c$, $\Omega_{\rm{MS}}$,  $\delta_{\rm{MS}}$, and $t_{\rm{MS}}$. We chose this parametrization of the Rabi frequencies because $\Omega_{\rm{MS}}^2 = \Omega_{\rm{MS(LI)}} \Omega_{\rm{MS (SI)}}$ is calibrated precisely, but we cannot assume $c=1$ because we do not precisely calibrate the SI and LI Rabi rates separately. We find good agreement between the data and fit. In principle it is possible to set $c>1$ deliberately if the SI laser intensity is too low to achieve $\Omega_{\rm{MS (SI)}}=\Omega_{\rm{MS (LI)}}$ and increasing the gate duration would lead to detrimental effects.
	These experiments confirm that the initial spin state of the SI does not significantly impact the spectroscopy signal of the LI and that the LI's signal faithfully represents the SI's spin resonance.
	
	When cooling the ions to the ground state and sweeping $\delta_{\rm{S}}$ in a Ramsey-type experiment with free evolution time $T_{\rm{R}}=\SI{1}{\milli\s}$, we observe the expected sinusoidal pattern for both species with a contrast in the LI signal of 0.89(1) and a phase offset between LI and SI oscillations of $\Delta\phi = 0.01(1)$ (Fig. \ref{fig:two-ion} (c)). The imperfect signal contrast, caused by experimental error sources (the leading contributions are most likely relative phase fluctuations between the Raman beam pairs), reduces the frequency sensitivity of the measurement (slope) by a factor equal to the contrast. This is not a fundamental limit of the method and could be improved by reducing the experimental imperfections. While the phase offset between SI and LI oscillations in this data is not significant, it is possible for a phase shift to arise from imperfect Ramsey-pulse calibration and finite temperature. This phase offset poses a potential issue for precision spectroscopy but simulations show that it can be suppressed by using a modified pulse sequence (see Appendix \ref{sec:phase_offset}).
	
	\subsection{Robustness of the method}
	To demonstrate the robustness of the multi-flip, we perform two sets of experiments under non-ideal conditions. In the first instance we investigate the effects of unwanted residual thermal motion. Instead of ground-state cooling, we only Doppler cool the LI of the ion pair (resulting in thermal states with mean occupations of $\nbar =  3.3(5)/1.7(1)$ for the axial in-phase/out-of-phase modes of motion; the MS interaction is again implemented using the in-phase mode). We tune the Rabi frequencies for both species to the optimal values for the respective $\bar{n}$. The LI shows a resonance with a contrast of $\approx 65\,\%$ in a Rabi experiment (Fig.~\ref{fig:two-ion} (d)) and $57(2)\,\%$ in a Ramsey experiment for the LI populations and a phase offset between the LI and SI oscillations of $\Delta\phi=0.21(4)$ (Fig.~\ref{fig:two-ion} (e)). Here, the reduction in contrast can be attributed to the relatively high LD-parameters of our experiment ($\eta_{\rm{Be}} \approx 0.17$ and $ \eta_{\rm{Mg}} \approx 0.29$ for the in-phase mode), which cause fluctuations in Debye-Waller factors. From simulations we find a contrast of $\approx 97\,\%$ for a Ramsey-type experiment using the same thermal excitation as in the experiment ($\nbar =  3.3$) but using LD-parameters of $\eta=0.1$ for LI and SI. The phase offset is expected from simulations, caused by the combination of large LD parameters, thermal occupation of the motion, and potential miscalibration of the pulse parameters. Simulations imply that a modified pulse sequence can be used to suppress the offset (see Appendix \ref{sec:phase_offset}).
	This ability to operate with imperfect motional ground state preparation in frequency measurement/clock applications can save time otherwise spent on ground-state cooling, improving the duty cycle (thereby reducing the Dick effect \cite{87Dick,98Santarelli}).
	The trade-off between improved duty cycle and reduction in contrast depends highly on the details of the experiment, e.g. initial Doppler temperature, heating rate of the trap, the ions’ LD-parameters, and other error sources on logic operations. One should also consider the effects of second order Doppler shifts when working at finite temperature. The target level of precision, together with these considerations, informs the preferred spectroscopy conditions as demonstrated, for example, across different aluminum ion clocks \cite{10Chou, 19Brewer}.
	
	
	In another set of experiments, we consider a case where $\Omega_{\mathrm{MS}}^2$ is not ideal. This could be useful in situations, where SI transition frequencies and corresponding coupling strengths are not well known and have to be found and identified first, as is the case with many molecular ions.
	As a test, we ground-state cool the ion pair, and set the Rabi frequency $\Omega_{\rm{MS(SI)}}$ of the SI MS interaction to roughly half (Fig. \ref{fig:two-ion} (f)) or twice the optimal value $\Omega_{\rm{MS}}$ (Fig. \ref{fig:two-ion} (g)). The experiment with the SI Rabi frequency reduced (increased) by a factor of two shows a contrast of $\approx 40\,\%$ ($\approx 70\,\%$) in the LI signal at $\delta_{\rm{S}}\approx0$. In addition to the reduction in contrast, the LI signal exhibits increased side-lobes when the SI is overdriven. 
	
	For $\Omega_{\mathrm{MS(SI)}}=2\Omega_{\mathrm{MS}}$, the additional resonances where the blue- (red-)detuned frequency component of the MS interaction is resonant with the blue (red) sideband transition ($\delta_{\rm{S}}=\pm \delta_{\rm{m}}$) are more strongly pronounced. Counter-intuitively, the SI signal shows a much stronger sideband resonance when it is under-driven (Fig. 1(f)). This is because the weakened MS interaction causes reduced coupling between the spins. When $\Omega_{\mathrm{SI}} > \Omega_{\mathrm{MS}}$, the MS interaction increases, causing the LI and SI spectra to overlap over a larger range of detunings.
	
	The capability to find transitions with an accuracy of $\approx \pm\delta_{\rm{MS}}$ combined with the LI's signal independence of the initial state of the SI and the possibility to operate with imperfect ground-state cooling could speed up searches for transitions. Once a transition is found, Rabi frequencies and transition frequencies can be determined more precisely.
	
	We fit all datasets as described in section \ref{sec:Implementation of basic method}. When fitting the Doppler-cooled data, the harmonic oscillator Hilbert space is truncated at $n = 20$ and a thermal state with $\nbar=3.3$ is used as the initial state.
	
	For comparison, simulations of traditional QLS using the thermal excitation measured in the experiment with just Doppler cooling give a contrast of $\approx23\,\%$ in a Ramsey sequence (compared to $\approx57\,\%$ with the multi-flip). The pulse durations are optimized numerically for best contrast. Both the opimized pulse durations and the contrast depend on the temperature. Lowering the LD-parameter as proposed for the multi-flip when using Doppler cooling results in a decrease of the contrast for the traditional QLS. For the above simulated example of $\nbar=3.3$ and LD-parameters $\eta=0.1$ for both the LI and SI, the traditional QLS would result in a contrast of $\approx18\,\%$.
	With the SI Rabi frequency reduced by a factor of two in a Rabi sequence using the traditional QLS we predict a contrast of 50\,\% (compared to $\approx40\,\%$ for the multi-flip). With the SI Rabi frequency increased by a factor of two in the traditional QLS no usable signal exists on resonance ($2\pi$ spin flip), but symmetric side peaks at $\approx \pm \,2\pi\times\SI{300}{\kilo\hertz}$ relative to the resonant frequency show a contrast of $\approx47\,\%$, which could also be used to identify the transition (compared to $\approx70\,\%$ on resonance for the multi-flip). We conclude that imperfect Rabi-frequency settings affect both the traditional QLS and multi-flip roughly equally.

	\subsection{Application to four ions}
\begin{figure}[b!]
		\resizebox{0.48 \textwidth}{!}{
			\includegraphics{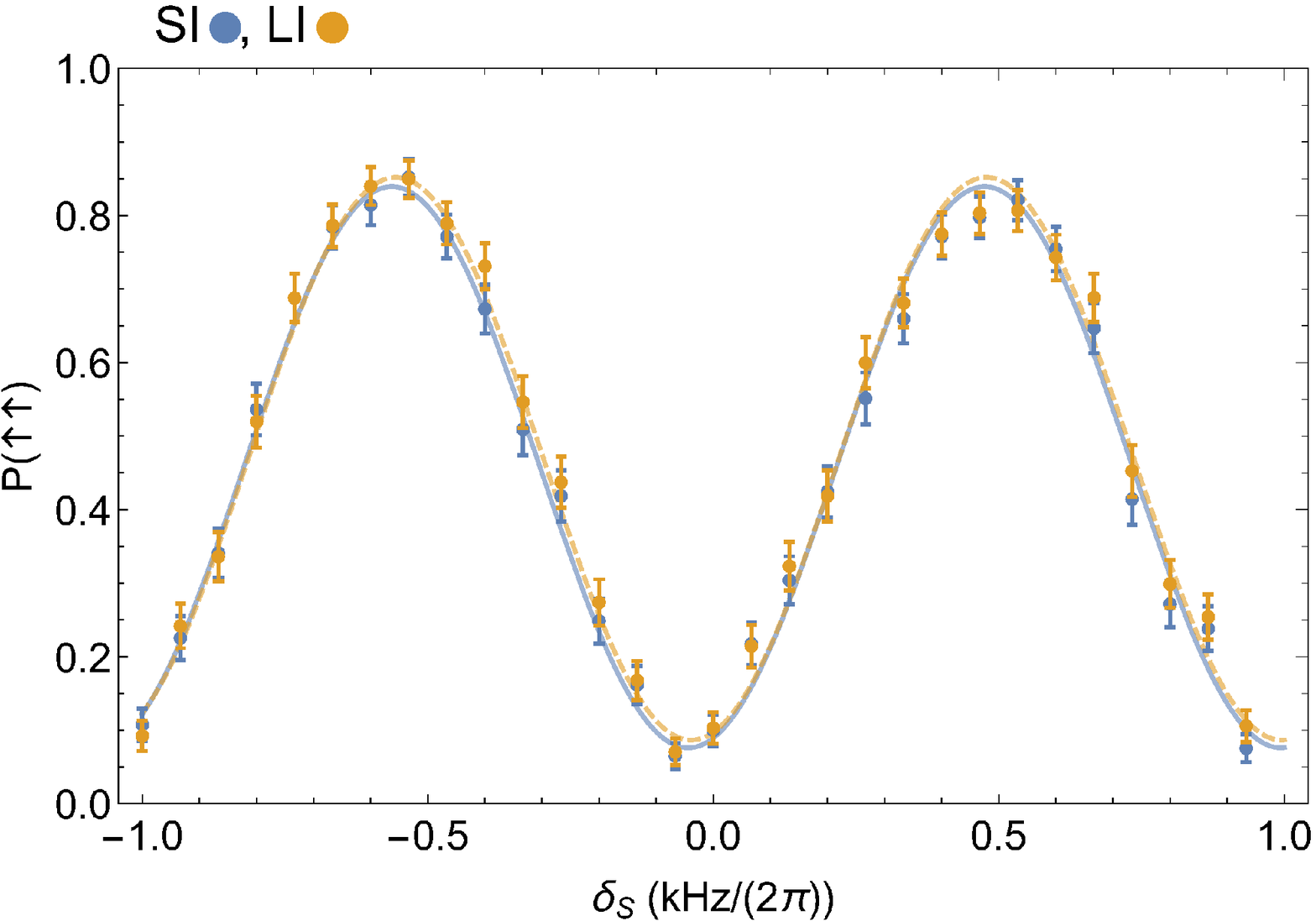}
		}
		\caption{Spin populations for the four-ion chain LI-SI-SI-LI (SI: blue, LI: orange), initially prepared in the spin state $\ket{\uparrow_{\rm{LI}} \uparrow_{\rm{SI}}  \uparrow_{\rm{SI}} \uparrow_{\rm{LI}}}$, performing Ramsey spectroscopy with free evolution time $T_{\rm{R}} = \SI{500}{\micro\s}$. The data shows the same Ramsey-fringe oscillation frequency as the LI-SI data with  $T_{\rm{R}} = \SI{1}{\milli\s}$, consistent with Heisenberg scaling. We fit sinusoids to the LI-LI and SI-SI data (displayed in matching colors with solid/dashed curves for fits to SI/LI data) and extract a contrast of 75(2)\,\% for the LI data and a phase shift between the LI-LI and SI-SI data of $\Delta\phi=0.03(3)$.
			In this experiment $t_{\rm{MS}} \approx \SI{64}{\micro\s}$. Error bars (1 sigma) are calculated assuming quantum projection noise only. Each data point represents the average outcome of 200 repetitions of the experiment.}
		\label{fig:four-ion}
	\end{figure}

	Finally, we demonstrate the behavior of the multi-flip for larger systems of ions in Ramsey-type spectroscopy.
	We use a four-ion linear chain of two SIs (\MgNoSpace) and two LIs (\BeNoSpace). The ions in the chain are ordered \BeNoSpace-\MgNoSpace-\MgNoSpace-\BeNoSpace. We implement the MS interaction on the motional mode in the axial direction, for which the \BeNoSpace-\Mg pair of the chain oscillates out of phase with the remaining \MgNoSpace-\Be pair. All axial motional modes are cooled to the ground state and the spin state of the ions is initialized to $\ket{\uparrow_{\rm{LI}} \uparrow_{\rm{SI}}  \uparrow_{\rm{SI}}  \uparrow_{\rm{LI}}}$. 
	We perform a Ramsey-type experiment (Fig. \ref{fig:four-ion}) with free evolution time $T_{\rm{R}}=\SI{500}{\micro\s}$ (Ramsey pulse duration \SI{64}{\micro\s}) and observe a contrast of 75(2)\,\% with the same Ramsey fringe oscillation frequency as the two-ion experiment for  $T_{\rm{R}}=\SI{1}{\milli\s}$, consistent with Heisenberg scaling. The LI-LI and SI-SI oscillations do not show a significant phase offset ($\Delta\phi = 0.03(3)$). A phase offset, which could again arise from imperfect calibration of the Ramsey pulses or other imperfections, can, as in the two-ion case, be canceled with a modified pulse sequence (see Appendix \ref{sec:phase_offset}).
	
	In practice, the signal to noise ratio is reduced below that expected for Heisenberg scaling due to loss in contrast from imperfect entangling operations \cite{11Monz}. The enhancement from multi-qubit entanglement depends on various aspects of the particular experimental realization, such as the properties of the local oscillator \cite{97Huelga, 04Andre}. In our experiment, the reduction in contrast is mainly caused by slow fluctuations of the relative phase between the Raman beams, which could be mitigated by actively stabilizing this phase. Other leading errors are caused by spontaneous emission and motional heating while mapping into and out of the GHZ state.
	
	\subsection{Applications in quantum information processing}
	The multi-flip can be used on resonance ($\delta_{\rm{S}}=0$) as a QND measurement method to detect if the SI (or qubit) is in the intended two-level manifold. In particular, if the SI can be described as a three-level system, with levels $a$ and $b$ defining a qubit and $c$ an auxiliary state, the multi-flip can be used to detect any leakage into level $c$, which in a quantum information processing context would be described as a \textit{leakage error}.
	When the multi-flip is applied to the $a$-$b$ state-manifold, the final state of the LI will indicate if a leakage event has occurred and if not, the qubit can be restored with a single-qubit rotation that reverses the rotation of the SI caused by the multi-flip. 
	Leakage will be relevant at some level for almost any quantum information or spectroscopy application. Streamlined QND detection of this error could reduce error-correction overhead \cite{18Brown}. Similar ideas about the use of the multi-flip for detecting leakage errors were contemplated by Schindler \textit{et al.} \cite{13Schindler}. 
	
	The multi-flip could also be used as a general QND readout method of the populations of a two-level system in trapped ions: invoking the three-level model described in the previous paragraph, the MS multi-flip could be used to determine the populations of the states $a$ and $b$ by applying the multi-flip to the state manifolds $a$-$c$ or $b$-$c$. This QND measurement could also be used to improve readout fidelity in frequency measurement and clock applications with potentially fewer repeated applications compared to \cite{07Koelemeij2,07Hume,11Hume} due to higher single-shot fidelity.

	\section{Conclusion}
	We have demonstrated a new method for QLS based on the MS interaction using a mixed-species ion system, showing its basic behavior, its robustness to experimental imperfections and how it scales in larger systems.
	This ``multi-flip'' technique provides a spectroscopy method with improved readout performance compared to traditional QLS. Additionally, it can  be used as a combined multi-ion-clock probe and readout technique, naturally providing Heisenberg scaling in Ramsey-type spectroscopy, which could help improve the stability of ion-based clocks. 
	Further applications of the multi-flip in quantum information include optimized QND error-syndrome measurements, which may help to scale future trapped-ion quantum computers to larger numbers of qubits. The MS multi-flip might also be generalized to other systems where oscillators can be coupled to two-level systems, such as superconducting circuits, the ro-vibrational levels of a molecule or micro/nanomechanical resonators.

	\paragraph{Acknowledgments}
	Authors D.K. and Y.W. have contributed equally to this work.
	The authors thank S. C. Burd and D. B. Hume for helpful comments on the manuscript. D.K., Y.W., and J.J.W.\ acknowledge support of the Professional Research Experience Program (PREP) operated jointly by NIST and University of Colorado Boulder. This work was supported by the Office of the Director of National Intelligence (ODNI) Intelligence Advanced Research Projects Activity (IARPA), and the NIST Quantum Information Program.
	D.K.\ acknowledges support from the Swiss National Science Foundation under grant no. 165208. S.D.E.\ acknowledges support by the U.S. National Science Foundation under Grant No. DGE 1650115.
		
	\appendix	
	
	\section{Methods}\label{sec:methods}
	
	\subsection{State preparation and readout}\label{sec:state_prep}
	At the start of each experiment, the \Be ion (LI) is optically pumped to \ket{2,2} in the S$_{1/2}$ electronic ground state.  Similarly, we optically pump the \Mg ion (SI) to the \ket{3,3} electronic ground state. 
	This is followed by Doppler cooling of \Mg and then \Be, implemented by driving the $\rm{S}_{1/2} \ket{3,3} \leftrightarrow \rm{P}_{3/2} \ket{4, 4}$ cycling-transition for \Mg and the $\rm{S}_{1/2} \ket{2,2} \leftrightarrow \rm{P}_{3/2} \ket{3, 3}$ cycling transition for \Be with $\sigma^+$-polarized light.
	For ground-state initialization of both axial modes of the \Be-\Mg ion pair, Raman sideband cooling is applied to the \Be ion after Doppler cooling \cite{95Monroe}. To transfer \Be to the $\ket{1, 1} = \ket{\uparrow_{\rm{Be}}}$ state we use a microwave composite pulse sequence that is robust against transition-frequency detuning and Rabi frequency errors. The sequence is composed of the following resonant pulses: $R(\pi,0)$, $R(\pi,5\pi/6)$, $R(\pi,\pi/3)$, $R(\pi,5\pi/6)$, $R(\pi, 0)$ \cite{86Levitt}, where $R(\theta,\phi)$ is the single-spin rotation defined as
	\begin{align}
	R(\theta,\phi)=
	\begin{pmatrix}
	\cos{(\theta/2)} & -i e^{-i\phi}\sin{(\theta/2)}\\
	-i e^{i\phi} \sin{(\theta/2)} &  \cos{(\theta/2)}
	\end{pmatrix}.
	\end{align} 
	With analogous pulse sequences, the \Mg \ket{3,3} state is transferred to the \ket{2,2} state and subsequently to the $\ket{3,1} = \ket{\uparrow_{\rm{Mg}}}$ state. 
	
	To measure the qubit states the initial mapping procedure is reversed, putting the $\ket{\uparrow_{\rm{Be}}}$, $\ket{\uparrow_{\rm{Mg}}}$ states back in their respective cycling transition ground states. The $\ket{\downarrow_{\rm{Be}}}\equiv \ket{2,0}$, $\ket{\downarrow_{\rm{Mg}}} \equiv \ket{2,0}$ states are shelved to \ket{1,-1} and \ket{2,-2}, respectively, using a single microwave $\pi$-pulse for the \Be $\ket{2,0}\rightarrow\ket{1,-1}$ transition and composite pulse sequences for the $\ket{2,0}\rightarrow\ket{3,-1}$ and $\ket{3,-1}\rightarrow\ket{2,-2}$ transitions of \Mg.
	We subsequently apply the Doppler-cooling laser beams tuned to resonance and collect the fluorescence light of the ions with an achromatic imaging system designed for both \SI{313}{\nano\m} and \SI{280}{\nano\m} light \cite{04Huang}. 
	With a detection duration of \SI{330}{\micro\s} for \Be (\SI{200}{\micro\s} for \Mg), we detect on average 30 photons per ion when they are in the \ket{2,2} (\ket{3,3}) state and 1.9 (0.7) photons when they are in the \ket{1,-1} (\ket{2,-2}) state (predominantly from light scattered by the trap structure). The qubit state is determined by choosing a photon-count threshold of 10 counts such that the states are maximally distinguished.
	To prepare the initial state of the four-ion string \Be-\Mg-\Mg-\Be we use the same techniques and ground-state cool all four axial modes. Here the qubit states are determined by fitting Poissionian distributions to histograms of the count data.
	
	\subsection{M\o lmer-S\o rensen interaction, four-ion entangled state}\label{sec:MS}
	The MS Hamiltonian in the interaction picture can be written (neglecting fast rotating terms, in the LD regime, and for a single motional mode) as
	\begin{align} \label{eq:H}
	\hat{H} = \hbar \, \sum_{j} \Omega_{\mathrm{MS},j} ( \hat{\sigma}^{+}_{j} \destroy \, e^{-i ((\delta_{\mathrm{S},j}-\delta_{\rm{m}}) t) + \phi_r} + \nonumber \\
	\hat{\sigma}^{+}_{j} \create e^{-i ((\delta_{\mathrm{S},j}+\delta_{\rm{m}}) t) + \phi_b} ) + H.c. \, ,
	\end{align}	
	where the sum runs over the number of ions with $j$ labeling the ion and $\Omega_{\rm{MS,j}}$ the Rabi frequency for each ion, defined as $\Omega_{\rm{MS,j}} = \eta_j \Omega_{\rm{0,MS,j}}$ with $\Omega_{\rm{0,MS,j}}$ the carrier Rabi frequency of each ion and $\eta_j = \delta k_j z_{0,j} b_j$ the respective ion's LD parameter. The difference of the $k$-vectors of the Raman beams addressing the $j$th ion is given by $\delta k_j$, $b_j$ is the normalized motional mode amplitude of the $j$th ion, and $z_{0,j}=\sqrt{\hbar/(2 m_j \omega_m)}$ with $m_j$ the mass of the $j$th ion and $\omega_m$ the oscillation frequency of the motional mode used. The detuning from the motional mode is $\delta_{\rm{m}}$, $t$ the duration, $\delta_{\rm{S, j}}$ the detuning from the respective ion's spin transition, $\hat{\sigma}^{+}_{j}$ the respective ion's spin raising operator, and \destroy (\create) the lowering (raising) operator of the motional mode used. Without loss of generality, we can set all optical phases to zero. Unless otherwise noted the Rabi frequencies are assumed to be equal for all ions ($\Omega_{\rm{MS,j}} = \Omega_{\rm{MS}}$).
	The implementation of the MS interaction on the two-ion pair is as described in \cite{15Tan} with the difference that the MS interaction is implemented on the axial in-phase mode of motion that has an oscillation frequency of $\approx 2\pi \times \SI{2.1}{\mega\hertz}$, resulting in LD parameters of $\eta_{\rm{Be}} \approx 0.17$ and $ \eta_{\rm{Mg}} \approx 0.29$ for the two species. We choose to implement the MS interaction for the four-ion chain (ordered \Be-\Mg-\Mg-\Be) on the motional mode in the axial direction, for which the \Be-\Mg part of the chain oscillates out of phase with the remaining \Mg-\Be part, which has an oscillation frequency of $\approx 2\pi \times \SI{3.1}{\mega\hertz}$, resulting in LD parameters of $|\eta_{\rm{Be}}| \approx 0.17$ and $|\eta_{\rm{Mg}}| \approx 0.14$ for the two species. When applying the MS interaction to the state $\ket{\uparrow_{\rm{Be}} \uparrow_{\rm{Mg}} \uparrow_{\rm{Mg}}  \uparrow_{\rm{Be}} }$ for a duration $t=2\pi/\delta_{\rm{MS}} \approx \SI{66}{\micro\s}$ we measure a state fidelity of 0.937(6) for the four-ion Schr\"odinger-cat type state $1 / \sqrt{2} \left( \ket{\uparrow_{\rm{Be}} \uparrow_{\rm{Mg}} \uparrow_{\rm{Mg}} \uparrow_{\rm{Be}} } \allowbreak + i  \ket{\downarrow_{\rm{Be}} \downarrow_{\rm{Mg}} \downarrow_{\rm{Mg}} \downarrow_{\rm{Be}} } \right)$, by performing an analysis of spin populations and spin parity measurements \cite{00SackettShort}. For this we measure the spin state of each ion individually by splitting the \Be-\Mg-\Mg-\Be ion chain into two \Be-\Mg ion pairs and detecting them sequentially by shuttling the pairs individually into the region where the detection laser beams are focused.
	
	\section{Example for dynamical decoupling of the LI} \label{sec:dyn_decoup}
	Simple pulse sequences for dynamical decoupling of the LI with $\pi$-pulses are given by the following.
	For Rabi-type spectroscopy: 
	$U_{\rm{MS}}(t_{\rm{MS}}/4), \,\allowbreak R_{\rm{LI}}(\pi,0), \,\allowbreak U_{\rm{MS}}(t_{\rm{MS}}/4), \,\allowbreak R_{\rm{LI}}(\pi,0), \,\allowbreak U_{\rm{MS}}(t_{\rm{MS}}/4), \,\allowbreak R_{\rm{LI}}(\pi,0), \,\allowbreak U_{\rm{MS}}(t_{\rm{MS}}/4)$, where $U_{\rm{MS}} (t)$ is the MS interaction applied for a duration $t$ and $R_{\rm{LI}}(\theta,\phi)$ 
	is a rotation of the LI spin as defined in Appendix \ref{sec:state_prep}.
	For Ramsey-type spectroscopy: $U_{\rm{MS}}(t_{\rm{MS}}/4), \,\allowbreak R_{\rm{LI}}(\pi,0), \,\allowbreak U_{\rm{MS}}(t_{\rm{MS}}/4), \,\allowbreak U_{\rm{free}}(T_{\rm{R}}/2), \,\allowbreak R_{\rm{LI}}(\pi,0),  \,\allowbreak U_{\rm{free}}(T_{\rm{R}}/2), \,\allowbreak U_{\rm{MS}}(t_{\rm{MS}}/4), \,\allowbreak R_{\rm{LI}}(\pi,0), \,\allowbreak U_{\rm{MS}}(t_{\rm{MS}}/4)$, where $U_{\rm{free}}(t)$ is the free evolution for the duration $t$. Using simulations, we find that a sequence with only a single $\pi$-pulse after half the Rabi or Ramsey sequence mirrors the detuned resonance feature, i.e.\ the resonance is centered on $-\delta_{\mathrm{S(LI)}}$ instead of $\delta_{\mathrm{S(LI)}}$ when the MS interaction is detuned by $\delta_{\mathrm{S(LI)}}$.
	
	\section{LI-SI Ramsey phase offset cancellation} \label{sec:phase_offset}
	Residual spin-motion entanglement after the MS interaction can lead to a phase offset in the Ramsey-type protocol. This can be caused by finite temperature, large LD parameters and imperfect calibration of MS-interaction parameters. The phase shift can be suppressed by increasing $T_{\rm R}$ or eliminated by constructing a four-measurement error signal instead of a two-measurement error signal \cite{89Morinaga}.
	The four-measurement error signal is given by
	\begin{align}
		e = & P_{\ket{\uparrow_{\rm{SI}}},\phi=\pi/2}(\uparrow_{\mathrm{LI}}) - P_{\ket{\uparrow_{\rm{SI}}},\phi=-\pi/2}(\uparrow_{\mathrm{LI}}) + \\
		& P_{\ket{\downarrow_{\rm{SI}}},\phi=\pi/2}(\uparrow_{\mathrm{LI}}) - P_{\ket{\downarrow_{\rm{SI}}},\phi=-\pi/2}(\uparrow_{\mathrm{LI}}), \nonumber 
	\end{align}
	where $P_{\ket{\uparrow_{\rm{SI}}}/\ket{\downarrow_{\rm{SI}}},\phi=\pm\pi/2}(\uparrow_{\mathrm{LI}})$ is the measured population of $\ket{\uparrow_{\mathrm{LI}}}$ after a Ramsey-type sequence with the initial SI state prepared to $\ket{\uparrow_{\rm{SI}}}/\ket{\downarrow_{\rm{SI}}}$ and the phases of the SI red and blue sideband tones shifted by $\pm \pi/2$ for the second Ramsey pulse relative to the first. This more complex protocol still features the same SNR as the original, i.e.\ a single round of the four-measurement sequence gives the same amount of information as two rounds of the two-measurement sequence.
	
	\section{Motional mode amplitudes for long ion chains} \label{sec:normal_mode_amplitudes}
	For high multi-flip contrast, the Rabi frequencies of all LIs and SIs should be the same. This can be challenging in long ion chains because the mode amplitudes will in general be different for different ions along the chain. However, this effect is least pronounced for the axial in-phase mode of motion. To demonstrate this, we calculate the mode amplitudes for \Ca,\Al chains with different numbers of ions and find that the axial in-phase mode amplitudes for ions of the same species differ by an experimentally insignificant amount. For example, in a symmetric ion chain with 7 \Al and 6 \Ca (alternating) ions confined by a potential with normal mode frequencies for a single \Al ion of $2 \pi \times \{6, 6.1, 1\}$~MHz (with the lowest value corresponding to the axial direction) we find relative differences in the axial in-phase mode amplitudes for each ion species to be below 1\%.

\end{document}